\pdfoutput=1
\documentclass[12pt,a4paper]{article}

\usepackage{ifthen} 
\newboolean{pdflatex}
\setboolean{pdflatex}{true} 

\newboolean{articletitles}
\setboolean{articletitles}{true} 

\newboolean{uprightparticles}
\setboolean{uprightparticles}{true} 

\newboolean{inbibliography}
\setboolean{inbibliography}{false} 

\def\paperauthors{LHCb collaboration} 
\def\paperasciititle{Observation of the~decay Lb->Psi(2S)PPi-} 
\def\papertitle{Observation of the~decay~\LbpsitwosPPi} 
\def\paperkeywords{{High Energy Physics}, {LHCb}} 
\def\papercopyright{\the\year\ CERN for the benefit of the LHCb collaboration} 
\def\paperlicence{CC-BY-4.0 licence}
\def\paperlicenceurl{https://creativecommons.org/licenses/by/4.0/}

\usepackage{dashrule} 
\usepackage{graphpap} 

\usepackage[top=1in, bottom=1.25in, left=1in, right=1in]{geometry}

\usepackage{microtype}
\usepackage{lineno}  
\usepackage{xspace} 
\usepackage{caption} 

\usepackage{graphicx}  
\usepackage{color}
\usepackage{colortbl}
\graphicspath{{./figs/}} 

\usepackage{amsmath} 
\usepackage{amssymb}
\usepackage{amsfonts}
\usepackage{upgreek} 

\usepackage[normalem]{ulem}

\newcommand*\patchAmsMathEnvironmentForLineno[1]{%
\expandafter\let\csname old#1\expandafter\endcsname\csname #1\endcsname
\expandafter\let\csname oldend#1\expandafter\endcsname\csname
end#1\endcsname
 \renewenvironment{#1}%
   {\linenomath\csname old#1\endcsname}%
   {\csname oldend#1\endcsname\endlinenomath}%
}
\newcommand*\patchBothAmsMathEnvironmentsForLineno[1]{%
  \patchAmsMathEnvironmentForLineno{#1}%
  \patchAmsMathEnvironmentForLineno{#1*}%
}
\AtBeginDocument{%
\patchBothAmsMathEnvironmentsForLineno{equation}%
\patchBothAmsMathEnvironmentsForLineno{align}%
\patchBothAmsMathEnvironmentsForLineno{flalign}%
\patchBothAmsMathEnvironmentsForLineno{alignat}%
\patchBothAmsMathEnvironmentsForLineno{gather}%
\patchBothAmsMathEnvironmentsForLineno{multline}%
\patchBothAmsMathEnvironmentsForLineno{eqnarray}%
}

\usepackage{hyperxmp}

\usepackage[pdftex,
            pdfauthor={\paperauthors},
            pdftitle={\paperasciititle},
            pdfkeywords={\paperkeywords},
            pdfcopyright={Copyright (C) \papercopyright},
            pdflicenseurl={\paperlicenceurl}]{hyperref}

\usepackage[all]{hypcap} 

\usepackage{xspace} 
\usepackage{upgreek}

\def\lhcb {\mbox{LHCb}\xspace}
\def\atlas  {\mbox{ATLAS}\xspace}

\def\MagUp {\mbox{\em Mag\kern -0.05em Up}\xspace}

\ifthenelse{\boolean{uprightparticles}}%
{
 
 \def\Pgamma      {\ensuremath{\upgamma}\xspace}

 \def\Pmu         {\ensuremath{\upmu}\xspace}

 \def\Ppi         {\ensuremath{\uppi}\xspace}

 \def\Pphi        {\ensuremath{\upphi}\xspace}                 
                  
 \def\Pchi        {\ensuremath{\upchi}\xspace}                 
 \def\Ppsi        {\ensuremath{\uppsi}\xspace}

 \def\PDelta      {\ensuremath{\Delta}\xspace}                 
 \def\PXi      {\ensuremath{\Xi}\xspace}                 
 \def\PLambda      {\ensuremath{\Lambda}\xspace}                 
 \def\PSigma      {\ensuremath{\Sigma}\xspace}                 
 \def\POmega      {\ensuremath{\Omega}\xspace}                 
 \def\PUpsilon      {\ensuremath{\Upsilon}\xspace}                 
 
 \def\PB      {\ensuremath{\mathrm{B}}\xspace}                 
                  
 \def\PD      {\ensuremath{\mathrm{D}}\xspace}

 \def\PJ      {\ensuremath{\mathrm{J}}\xspace}                 
 \def\PK      {\ensuremath{\mathrm{K}}\xspace}

 \def\PS      {\ensuremath{\mathrm{S}}\xspace}

 \def\PX      {\ensuremath{\mathrm{X}}\xspace}                 
                  
 \def\PZ      {\ensuremath{\mathrm{Z}}\xspace}                 
                  
 \def\Pb      {\ensuremath{\mathrm{b}}\xspace}                 
 \def\Pc      {\ensuremath{\mathrm{c}}\xspace}

 \def\Pi      {\ensuremath{\mathrm{i}}\xspace}

 \def\Pp      {\ensuremath{\mathrm{p}}\xspace}

 \def\Ps      {\ensuremath{\mathrm{s}}\xspace}

}
{
 
 \def\Pgamma      {\ensuremath{\gamma}\xspace}

 \def\Pmu         {\ensuremath{\mu}\xspace}

 \def\Ppi         {\ensuremath{\pi}\xspace}

 \def\Pphi        {\ensuremath{\phi}\xspace}                 
                  
 \def\Pchi        {\ensuremath{\chi}\xspace}                 
 \def\Ppsi        {\ensuremath{\psi}\xspace}                 
                  
 \mathchardef\PDelta="7101
 \mathchardef\PXi="7104
 \mathchardef\PLambda="7103
 \mathchardef\PSigma="7106
 \mathchardef\POmega="710A
 \mathchardef\PUpsilon="7107
                  
 \def\PB      {\ensuremath{B}\xspace}                 
                  
 \def\PD      {\ensuremath{D}\xspace}

 \def\PJ      {\ensuremath{J}\xspace}                 
 \def\PK      {\ensuremath{K}\xspace}

 \def\PS      {\ensuremath{S}\xspace}

 \def\PX      {\ensuremath{X}\xspace}                 
                  
 \def\PZ      {\ensuremath{Z}\xspace}                 
                  
 \def\Pb      {\ensuremath{b}\xspace}                 
 \def\Pc      {\ensuremath{c}\xspace}

 \def\Pi      {\ensuremath{i}\xspace}

 \def\Pp      {\ensuremath{p}\xspace}

 \def\Ps      {\ensuremath{s}\xspace}

}

\makeatletter
\ifcase \@ptsize \relax
  \newcommand{\miniscule}{\@setfontsize\miniscule{4}{5}}
\or
  \newcommand{\miniscule}{\@setfontsize\miniscule{5}{6}}
\or
  \newcommand{\miniscule}{\@setfontsize\miniscule{5}{6}}
\fi
\makeatother

\DeclareRobustCommand{\optbar}[1]{\shortstack{{\miniscule (\rule[.5ex]{1.25em}{.18mm})}
  \\ [-.7ex] $#1$}}

\def\mup        {{\ensuremath{\Pmu^+}}\xspace}
\def\mun        {{\ensuremath{\Pmu^-}}\xspace} 
\def\mumu       {{\ensuremath{\Pmu^+\Pmu^-}}\xspace}

\def\g      {{\ensuremath{\Pgamma}}\xspace}

\def\Z      {{\ensuremath{\PZ}}\xspace}

\def\squark    {{\ensuremath{\Ps}}\xspace}

\def\cquark    {{\ensuremath{\Pc}}\xspace}

\def\bquark    {{\ensuremath{\Pb}}\xspace}
\def\bquarkbar {{\ensuremath{\overline \bquark}}\xspace}

\def\pion   {{\ensuremath{\Ppi}}\xspace}

\def\pip    {{\ensuremath{\pion^+}}\xspace}
\def\pim    {{\ensuremath{\pion^-}}\xspace}

\def\kaon    {{\ensuremath{\PK}}\xspace}
  \def\Kbar    {{\kern 0.2em\overline{\kern -0.2em \PK}{}}\xspace}

\def\KorKbar    {\kern 0.18em\optbar{\kern -0.18em K}{}\xspace}

\def\Kp      {{\ensuremath{\kaon^+}}\xspace}
\def\Km      {{\ensuremath{\kaon^-}}\xspace}

\def\KS      {{\ensuremath{\kaon^0_{\mathrm{ \scriptscriptstyle S}}}}\xspace}

  \def\Dbar    {{\kern 0.2em\overline{\kern -0.2em \PD}{}}\xspace}
\def\D       {{\ensuremath{\PD}}\xspace}

\def\DorDbar    {\kern 0.18em\optbar{\kern -0.18em D}{}\xspace}
\def\Dz      {{\ensuremath{\D^0}}\xspace}

\def\Dstarp  {{\ensuremath{\D^{*+}}}\xspace}

\def\Ds      {{\ensuremath{\D^+_\squark}}\xspace}

\def\B       {{\ensuremath{\PB}}\xspace}
\def\Bbar    {{\ensuremath{\kern 0.18em\overline{\kern -0.18em \PB}{}}}\xspace}

\def\BorBbar    {\kern 0.18em\optbar{\kern -0.18em B}{}\xspace}

\def\Bu      {{\ensuremath{\B^+}}\xspace}

\def\Bp      {{\ensuremath{\Bu}}\xspace}

\def\Bd      {{\ensuremath{\B^0}}\xspace}
\def\Bs      {{\ensuremath{\B^0_\squark}}\xspace}

\def\jpsi     {{\ensuremath{{\PJ\mskip -3mu/\mskip -2mu\Ppsi\mskip 2mu}}}\xspace}
\def\psitwos  {{\ensuremath{\Ppsi{(2\PS)}}}\xspace}

  \def\Y#1S{\ensuremath{\PUpsilon{(#1S)}}\xspace}

\def\proton      {{\ensuremath{\Pp}}\xspace}

\def\Lz          {{\ensuremath{\PLambda}}\xspace}
\def\Lbar        {{\ensuremath{\kern 0.1em\overline{\kern -0.1em\PLambda}}}\xspace}
\def\LorLbar    {\kern 0.18em\optbar{\kern -0.18em \PLambda}{}\xspace}

\def\Lb      {{\ensuremath{\Lz^0_\bquark}}\xspace}

\def\Lc      {{\ensuremath{\Lz^+_\cquark}}\xspace}

\def\BF         {{\ensuremath{\mathcal{B}}}\xspace}

\def\BR         {\BF}
\newcommand{\decay}[2]{\ensuremath{#1\!\to #2}\xspace}         

\def\to                 {\ensuremath{\rightarrow}\xspace}

\def\AT#1     {\ensuremath{A_{\mathrm{T}}^{#1}}\xspace}           

\def\C#1      {\ensuremath{\mathcal{C}_{#1}}\xspace}                       
\def\Cp#1     {\ensuremath{\mathcal{C}_{#1}^{'}}\xspace}                    
\def\Ceff#1   {\ensuremath{\mathcal{C}_{#1}^{\mathrm{(eff)}}}\xspace}        
\def\Cpeff#1  {\ensuremath{\mathcal{C}_{#1}^{'\mathrm{(eff)}}}\xspace}       
\def\Ope#1    {\ensuremath{\mathcal{O}_{#1}}\xspace}                       
\def\Opep#1   {\ensuremath{\mathcal{O}_{#1}^{'}}\xspace}                    

\newcommand{\tev}{\ifthenelse{\boolean{inbibliography}}{\ensuremath{~T\kern -0.05em eV}}{\ensuremath{\mathrm{\,Te\kern -0.1em V}}}\xspace}
\newcommand{\gev}{\ensuremath{\mathrm{\,Ge\kern -0.1em V}}\xspace}
\newcommand{\mev}{\ensuremath{\mathrm{\,Me\kern -0.1em V}}\xspace}
\newcommand{\kev}{\ensuremath{\mathrm{\,ke\kern -0.1em V}}\xspace}
\newcommand{\ev}{\ensuremath{\mathrm{\,e\kern -0.1em V}}\xspace}
\newcommand{\gevc}{\ensuremath{{\mathrm{\,Ge\kern -0.1em V\!/}c}}\xspace}
\newcommand{\mevc}{\ensuremath{{\mathrm{\,Me\kern -0.1em V\!/}c}}\xspace}
\newcommand{\gevcc}{\ensuremath{{\mathrm{\,Ge\kern -0.1em V\!/}c^2}}\xspace}
\newcommand{\gevgevcccc}{\ensuremath{{\mathrm{\,Ge\kern -0.1em V^2\!/}c^4}}\xspace}
\newcommand{\mevcc}{\ensuremath{{\mathrm{\,Me\kern -0.1em V\!/}c^2}}\xspace}

\def\mm   {\ensuremath{\mathrm{ \,mm}}\xspace}

\def\mum  {\ensuremath{{\,\upmu\mathrm{m}}}\xspace}

\def\invfb   {\ensuremath{\mbox{\,fb}^{-1}}\xspace}

\newcommand{\chisq}{\ensuremath{\chi^2}\xspace}

\newcommand{\chisqip}{\ensuremath{\chi^2_{\text{IP}}}\xspace}

\def\gsim{{~\raise.15em\hbox{$>$}\kern-.85em
          \lower.35em\hbox{$\sim$}~}\xspace}
\def\lsim{{~\raise.15em\hbox{$<$}\kern-.85em
          \lower.35em\hbox{$\sim$}~}\xspace}

\def\sPlot{\mbox{\em sPlot}\xspace}

\def\pt         {\mbox{$p_{\mathrm{ T}}$}\xspace}

\def\evtgen     {\mbox{\textsc{EvtGen}}\xspace}

\def\geant      {\mbox{\textsc{Geant4}}\xspace}

\def\photos     {\mbox{\textsc{Photos}}\xspace}

\def\pythia     {\mbox{\textsc{Pythia}}\xspace}

\def\tell1  {TELL1\xspace}
\def\ukl1   {UKL1\xspace}

\newcommand{\eg}{\mbox{\itshape e.g.}\xspace}

\usepackage{rotating}

\def\psitwosmumu{\decay{\psitwos}{\mumu}}
\def\LbjpsiPPi		{\decay{\Lb}{\jpsi	 \Pp \pim					}}
\def\LbjpsiPK		{\decay{\Lb}{\jpsi	 \Pp \Km 					}}
\def\LbpsitwosPPi	{\decay{\Lb}{\psitwos\Pp \pim					}}
\def\LbpsitwosPK	{\decay{\Lb}{\psitwos\Pp \Km 					}}

\def\LbpsitwosLz	{$\decay{\Lb}{\psitwos\Lz						}$}

\def\psitwosPPi	{\ensuremath{\psitwos \Pp \pim}}
\def\psitwosPK	{\ensuremath{\psitwos \Pp \Km }}
\def\psitwosP	{\ensuremath{\psitwos \Pp  }}
\def\psitwosPi	{\ensuremath{\psitwos \pim }}
\def\PPi    	{\ensuremath{\Pp      \pim }}
\def\LbchicPK	        {\decay{\Lb}{\Pchi_{\cquark}\Pp\Km}}

\def\sigmastud		{\ensuremath{5.23 \pm 0.55}}
\def\sigmanorm		{\ensuremath{3.96 \pm 0.13}}
\def\yieldstud		{\ensuremath{121 \pm 13}}
\def\yieldnorm		{\ensuremath{806 \pm 29}}
\def\effratio		{\ensuremath{0.761 \pm 0.004}}
\def\brratio		{\ensuremath{\left(11.4 \pm 1.3 \pm 0.2\right)\!\%}}

\def\systFitModel	{\ensuremath{  0.7  }}

\def\systPID		{\ensuremath{  0.2  }}
\def\systDaSiAgr	{\ensuremath{  1.0  }}
\def\systTrigger	{\ensuremath{  1.1  }}
\def\systSSS        {\ensuremath{  0.5  }}
\def\systTotal		{\ensuremath{  1.7  }}

\usepackage{cite} 
\usepackage{mciteplus}

\usepackage{longtable} 

\begin{document}

\renewcommand{\thefootnote}{\fnsymbol{footnote}}
\setcounter{footnote}{1}


\begin{titlepage}
\pagenumbering{roman}

\vspace*{-1.5cm}
\centerline{\large EUROPEAN ORGANIZATION FOR NUCLEAR RESEARCH (CERN)}
\vspace*{1.5cm}
\noindent
\begin{tabular*}{\linewidth}{lc@{\extracolsep{\fill}}r@{\extracolsep{0pt}}}
\ifthenelse{\boolean{pdflatex}}
{\vspace*{-1.5cm}\mbox{\!\!\!\includegraphics[width=.14\textwidth]{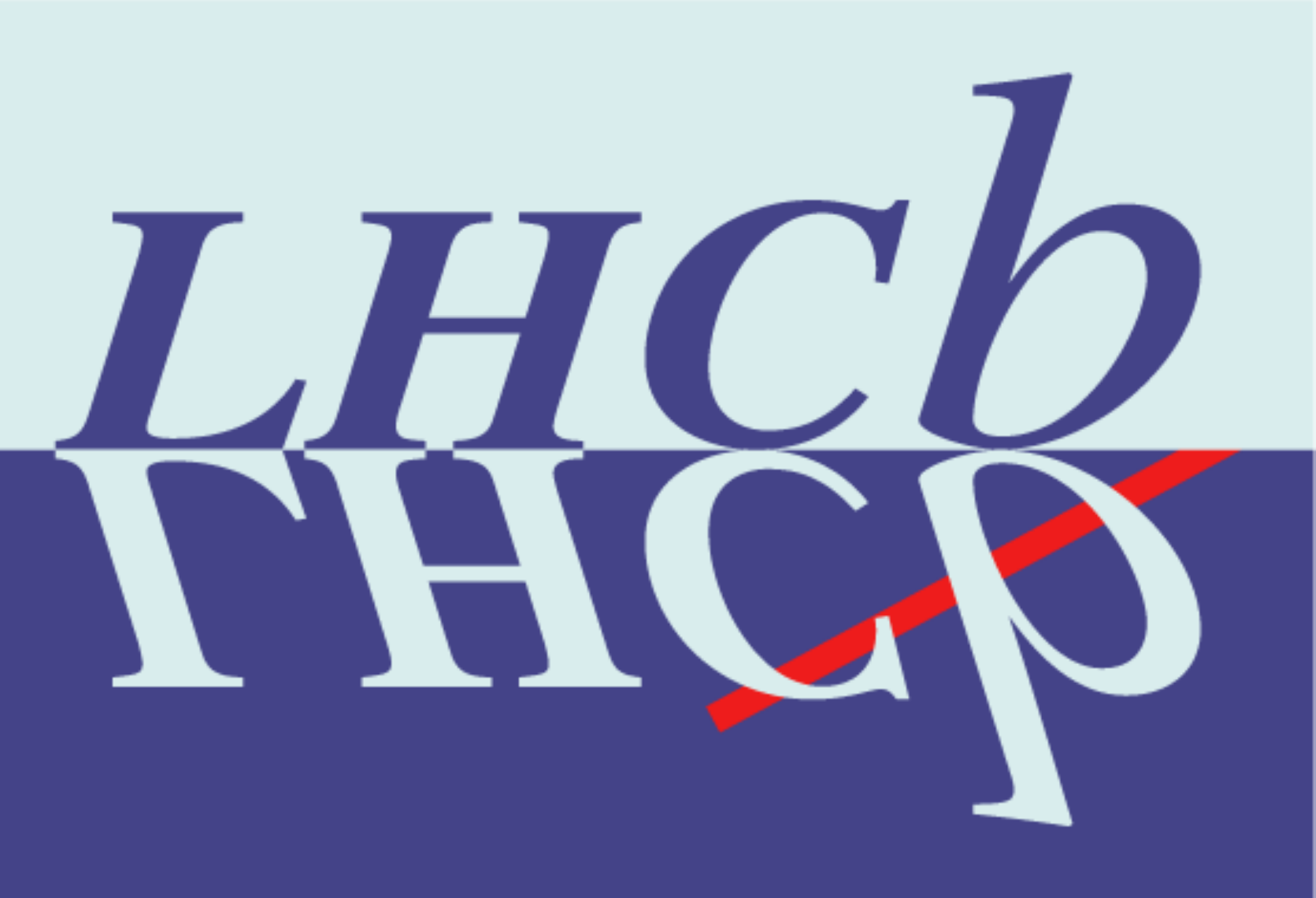}} & &}%
{\vspace*{-1.2cm}\mbox{\!\!\!\includegraphics[width=.12\textwidth]{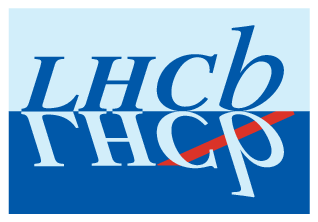}} & &}%
\\
& & CERN-EP-2018-156 \\  
& & LHCb-PAPER-2018-022 \\  
& & June 20, 2018
\end{tabular*}

\vspace*{4.0cm}

{\normalfont\bfseries\boldmath\huge
\begin{center}
  \papertitle 
\end{center}
}

\vspace*{1.5cm}

\begin{center}
\paperauthors\footnote{Authors are listed at the end of this paper.}
\end{center}

\vspace{\fill}

\begin{abstract}
  \noindent
  The~Cabibbo-suppressed decay \LbpsitwosPPi is observed for the
  first time using a~data sample 
  collected by the~\lhcb
  experiment in proton\nobreakdash-proton collisions 
  corresponding to 1.0, 2.0 and 1.9\invfb of  integrated
  luminosity at 
  centre\nobreakdash-of\nobreakdash-mass
  energies of 7, 8 and 13\tev, respectively.
  The~\psitwos~mesons are reconstructed in the~$\mup\mun$ final state.
  The~branching fraction with respect to that of 
  the~\mbox{\LbpsitwosPK}~decay mode is measured to be
  \begin{equation*}
    \dfrac{\BR\left( \Lb\to\psitwos\proton\pim\right)}{\BR\left( \Lb\to\psitwos\proton\Km\right)}=\left(11.4 \pm 1.3 \pm 0.2\right)\!\%\,,
\end{equation*}
  where the~first uncertainty is statistical and the~second is systematic.
  The~$\psitwos\proton$ and $\psitwos\pim$~mass spectra are investigated
  and no~evidence for exotic resonances is found. 
\end{abstract}

\vspace*{1.5cm}

\begin{center}
  Published in JHEP 1808\,(2018) 131
\end{center}

\vspace{\fill}
{\footnotesize 
\centerline{\copyright~\papercopyright. \href{\paperlicenceurl}{\paperlicence}.}}
\vspace*{2mm}

\end{titlepage}


\newpage
\setcounter{page}{2}
\mbox{~}
%

\cleardoublepage


\renewcommand{\thefootnote}{\arabic{footnote}}
\setcounter{footnote}{0}



\pagestyle{plain} 
\setcounter{page}{1}
\pagenumbering{arabic}


%


\section{Introduction}
\label{sec:Introduction}

The~\Lb~baryon is the~isospin\nobreakdash-singlet ground state 
of a~bound system of a~beauty quark and two light quarks.
The~high production rate of \bquark~quarks 
at the~Large Hadron Collider\,(LHC)~\cite{LHCb-PAPER-2010-002,
  LHCb-PAPER-2011-003,
  LHCb-PAPER-2013-016,
  LHCb-PAPER-2015-037,
  LHCb-PAPER-2016-031},
along with the~excellent mass resolution and hadron\nobreakdash-identification capabilities 
of the~LHCb detector, give access to a~variety of decay channels 
of the~\Lb~baryon, 
including 
multibody, 
rare, 
charmless and semileptonic decays~\cite{LHCb-PAPER-2011-016,
  LHCb-PAPER-2012-057,
  LHCb-PAPER-2013-025,
  LHCb-PAPER-2013-056,
  LHCb-PAPER-2013-061,
  LHCb-PAPER-2014-002,
  LHCb-PAPER-2014-004,
  LHCb-PAPER-2014-020,
  LHCb-PAPER-2015-009,
  LHCb-PAPER-2015-032,
  LHCb-PAPER-2015-060,
  LHCb-PAPER-2016-002,
  LHCb-PAPER-2016-004,
  LHCb-PAPER-2016-049,
  LHCb-PAPER-2016-059,
  LHCb-PAPER-2016-061,
  LHCb-PAPER-2017-011,
  LHCb-PAPER-2017-016,
  LHCb-PAPER-2017-034,
  LHCb-PAPER-2018-005}.
The~high signal yield of the~\mbox{$\Lb\to\jpsi\proton\Km$}~decay~\cite{LHCb-PAPER-2015-032}
facilitated a~precise measurement of 
the~\Lb lifetime~\cite{LHCb-PAPER-2014-003}, 
while the~relatively low energy released in 
the~\mbox{\LbpsitwosPK}  and~\mbox{\LbchicPK}~decays allowed for
precise measurements of the~\Lb~mass~\cite{LHCb-PAPER-2015-060,LHCb-PAPER-2017-011}.
A~six\nobreakdash-dimensional  amplitude analysis 
of the~\mbox{$\Lb\to\jpsi\proton\Km$}~decay resulted 
in the~observation of the~$\mathrm{P}_{\cquark}(4380)^{+}$~and 
$\mathrm{P}_{\cquark}(4450)^{+}$~pentaquark states decaying 
into~the~$\jpsi\proton$~final state\cite{LHCb-PAPER-2015-029}.
Later, these states were confirmed using a~model\nobreakdash-independent
technique~\cite{LHCb-PAPER-2016-009}.
Subsequently, an~analysis of Cabibbo\nobreakdash-suppressed
\LbjpsiPPi~decays found evidence for contributions from 
the~$\mathrm{P}_{\cquark}(4380)^{+}$~and~$\mathrm{P}_{\cquark}(4450)^{+}$~pentaquarks
and from the~$\Z_{\cquark}(4200)^{-}$~tetraquark~\cite{LHCb-PAPER-2016-015}.

The~first~observation of \Lb~decays to the~excited charmonium state \psitwos~was made
in the~\mbox{\LbpsitwosLz}~decay mode by the~\atlas~collaboration~\cite{Aad:2015msa}.
Later, the~decay~\mbox{\LbpsitwosPK} was observed by the~LHCb 
collaboration~\cite{LHCb-PAPER-2015-060}.
The~Cabibbo\nobreakdash-suppressed analogue of the~latter decay, 
\mbox{$\Lb\to\psitwos\proton\pim$},
is of particular interest because of possible 
contributions from exotic states in both  
the~$\psitwos\proton$~system, similar to the~$\mathrm{P}_{\cquark}(4380)^{+}$~and 
$\mathrm{P}_{\cquark}(4450)^{+}$~pentaquark states, 
and in the~$\psitwos\pim$~system,
analogous to the~charged charmonium\nobreakdash-like state~$\Z_{\cquark}(4430)^-$
studied in detail by the~Belle and LHCb collaborations 
in \mbox{$\B\to\psitwos\pim\kaon$}~decays~\cite{Choi:2007wga,
  Mizuk:2009da,Chilikin:2013tch,
  LHCb-PAPER-2014-014,
  LHCb-PAPER-2015-038}.
Depending on the~nature of a~proposed exotic state, its coupling
with the~\psitwos~meson can be larger than with the~\jpsi~meson.
For~example, the~decay rate of the~$\PX(3872)$~particle to 
the~$\psitwos\g$~final state was found to exceed the~corresponding decay rate
to the~\mbox{$\jpsi\g$}~final state~\cite{Aubert:2008ae,LHCb-PAPER-2014-008}.

This~paper reports the~first observation of the~decay~\mbox{\LbpsitwosPPi}
using a~data sample 
collected by the~\lhcb
experiment in proton\nobreakdash-proton collisions 
corresponding to 1.0, 2.0 and 1.9\invfb of  integrated
luminosity at 
centre\nobreakdash-of\nobreakdash-mass
energies of 7, 8 and 13\tev, respectively.
A~measurement is made of 
the~~\mbox{\LbpsitwosPPi}~branching fraction 
relative to that of the~Cabibbo\nobreakdash-favoured decay~\mbox{\LbpsitwosPK},
\begin{equation}
  R_{\pion/\kaon}\equiv\dfrac{\BR\left(\LbpsitwosPPi\right)}{\BR\left(\LbpsitwosPK\right)}\,,
  \label{eq:br_ratio}
\end{equation}
where the~\psitwos~mesons are reconstructed
in the~\mumu~final state. Throughout this~paper
the~inclusion of charge\nobreakdash-conjugated processes
is implied.

\section{Detector and simulation}
\label{sec:Detector}

The~\lhcb detector~\cite{Alves:2008zz,LHCb-DP-2014-002} is a~single\nobreakdash-arm forward
spectrometer covering the~\mbox{pseudorapidity} range~\mbox{$2<\eta<5$},
designed for the~study of particles containing~\bquark~or~\cquark~quarks.
The~detector includes a~high-precision tracking system
consisting of a~silicon\nobreakdash-strip vertex detector surrounding
the~$\Pp\Pp$~interaction region,
a~large\nobreakdash-area~silicon\nobreakdash-strip detector located
upstream of a~dipole magnet with a~bending power of about
$4{\mathrm{\,Tm}}$, and three stations of silicon\nobreakdash-strip detectors and straw
drift tubes
placed downstream of the~magnet.
The~tracking system provides a~measurement of the~momentum 
of charged particles with
a~relative uncertainty that varies from 0.5\% at low momentum to 1.0\% at 200\gevc.
The~minimum distance of a~track to a~primary vertex~(PV), the~impact parameter~(IP), 
is measured with a~resolution of $(15+29/\pt)\mum$,
where~\pt~is the~component of the~momentum transverse to the~beam, in\,\gevc.
Different types of charged hadrons are distinguished using information
from two ring\nobreakdash-imaging Cherenkov detectors~(RICH).
Photons, electrons and hadrons are identified by a~calorimeter system consisting of
scintillating\nobreakdash-pad and preshower detectors, an~electromagnetic
calorimeter and a~hadronic calorimeter. Muons are identified by a~system
composed of alternating layers of iron and multiwire
proportional chambers.

The~online event selection is performed by a~trigger~\cite{LHCb-DP-2012-004},
which consists of a~hardware stage, based on information from the~calorimeter and
muon systems, followed by a~software stage, which applies a~full event reconstruction.
The~hardware trigger selects muon candidates with high transverse momentum or
dimuon candidates  with high value of the~product of the~\pt of each muon.
The~subsequent software trigger is composed of two stages, the first of which performs
a~partial event reconstruction, while full event reconstruction is done at the~second stage.
In the~software trigger, each~pair of oppositely
charged muons forming a~good\nobreakdash-quality two\nobreakdash-track vertex 
is required to be significantly displaced from all PVs and 
the~mass of the~pair is required to exceed~$2.7\gevcc$.

The~techniques used in this analysis are validated using simulated events.
In~the~simulation, $\Pp\Pp$~collisions are generated using
\pythia~\cite{Sjostrand:2007gs} 
 with a~specific \lhcb configuration~\cite{LHCb-PROC-2010-056}.
Decays of hadronic particles are described by \evtgen~\cite{Lange:2001uf},
in which final\nobreakdash-state radiation is generated
using \photos~\cite{Golonka:2005pn}.
The~interaction of the~generated particles with the~detector,
and its response, are implemented using the~\geant
toolkit~\cite{Allison:2006ve, *Agostinelli:2002hh} as described
in Ref.~\cite{LHCb-PROC-2011-006}.

\section{Event selection}
\label{sec:Selection}

The~signal~\LbpsitwosPPi~and the~normalization~\LbpsitwosPK~decays are 
both reconstructed using the~decay mode~\psitwosmumu. 
Similar selection criteria, based on those used in
Ref.~\cite{LHCb-PAPER-2015-060}, are applied to both channels.


Muon, proton, pion and kaon candidates are identified
using combined information from the~RICH, calorimeter and muon detectors.
They~are required to have a~transverse momentum larger
than $550$, $900$, $500$ and $200\mevc$, respectively.
To~allow for an~efficient particle identification, 
kaons and pions are required to~have a~momentum
between 3.2~and 150\gevc, whilst protons must have a~momentum
between 10~and 150\gevc.
To reduce the~combinatorial background due to particles produced in the~$\proton\proton$~interaction,
only tracks that are inconsistent with originating from a~PV are used.

Pairs of oppositely charged muons consistent with originating from a~common vertex are combined to
form \psitwosmumu~candidates. The~mass of the~dimuon candidate is required to be
between $3.67$ and $3.70\gevcc$, where the~asymmetric mass range around
the~known $\psitwos$~mass~\cite{PDG2018} is chosen to 
account  for final\nobreakdash-state radiation.
The~position of the~reconstructed dimuon vertex is 
required to be inconsistent with that of any of 
the~reconstructed PVs. 

To~form signal\,(normalization) \Lb~candidates, 
the~selected~\psitwos~candidates are 
combined with a~proton and a~pion\,(kaon) of opposite charges.
Each~\Lb~candidate is associated with the~PV with 
respect to which it has the~smallest~\chisqip,
where \chisqip\ is defined as the~difference in the~vertex-fit \chisq of
a~given PV reconstructed with and without the~particle under consideration.
To~improve the~\Lb~mass resolution, a~kinematic fit~\cite{Hulsbergen:2005pu}
is performed. This~fit constrains 
the~four charged final\nobreakdash-state particles to form common vertex, 
the~mass of the~\mumu~combination to the~known 
\psitwos~mass and the~\Lb~candidate to originate from the~associated~PV.
A~good quality of this~fit is required to further suppress combinatorial background.
In~addition, the~measured decay time of the~\Lb~candidate, calculated with respect
to the~associated PV, is required to be
between $0.2$ and $2.0\mm/c$ to suppress 
poorly reconstructed candidates and background
from particles originating from the~PV.

To suppress cross\nobreakdash-feed from $\Bd\to\psitwos\Kp\pim$~decays 
with the~positively charged
kaon\,(negatively charged pion)
misidentified as a~proton\,(antiproton) for the~signal\,(normalization) channel, 
a~veto is applied on the~\Lb~candidate mass
recalculated with a~kaon\,(pion) mass hypothesis for the~proton.
Any~candidate~with a~recalculated mass consistent with the~nominal \Bd~mass is rejected.
A~similar veto is applied to suppress cross\nobreakdash-feed 
from~\mbox{$\Bs\to\psitwos\Km\Kp$}~decays
with the~positively charged kaon
misidentified as a~proton, and additionally for  the~signal channel,
the~negatively charged kaon misidentified as a~pion.
Finally, to suppress cross\nobreakdash-feed from 
the~\mbox{$\decay{\Lb}{\psitwos\Lz}$}~decay,  
followed  by a~\mbox{\decay{\Lz}{\proton\pim}}~decay,
candidates~with a~$\proton\pim$~mass that  is consistent
with the~nominal $\Lz$~mass~\cite{PDG2018} are rejected.

\section{Signal yields and efficiencies}
\label{sec:sig_eff}

\begin{figure}[tb]
  \setlength{\unitlength}{1mm}
  \centering
  \begin{picture}(150,60)
    \put( 0,0){\includegraphics*[width=75mm,height=60mm]{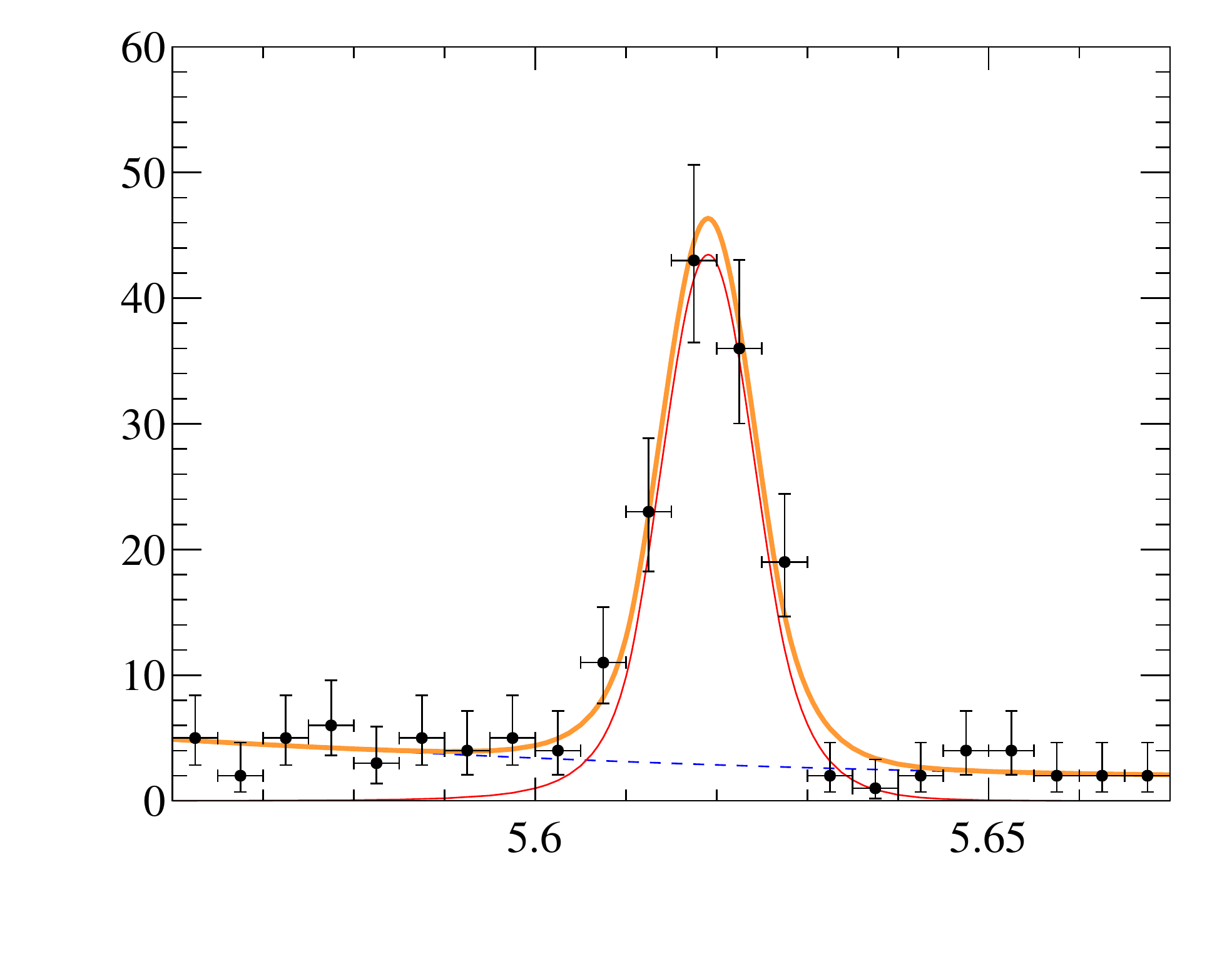}}
    \put(75,0){\includegraphics*[width=75mm,height=60mm]{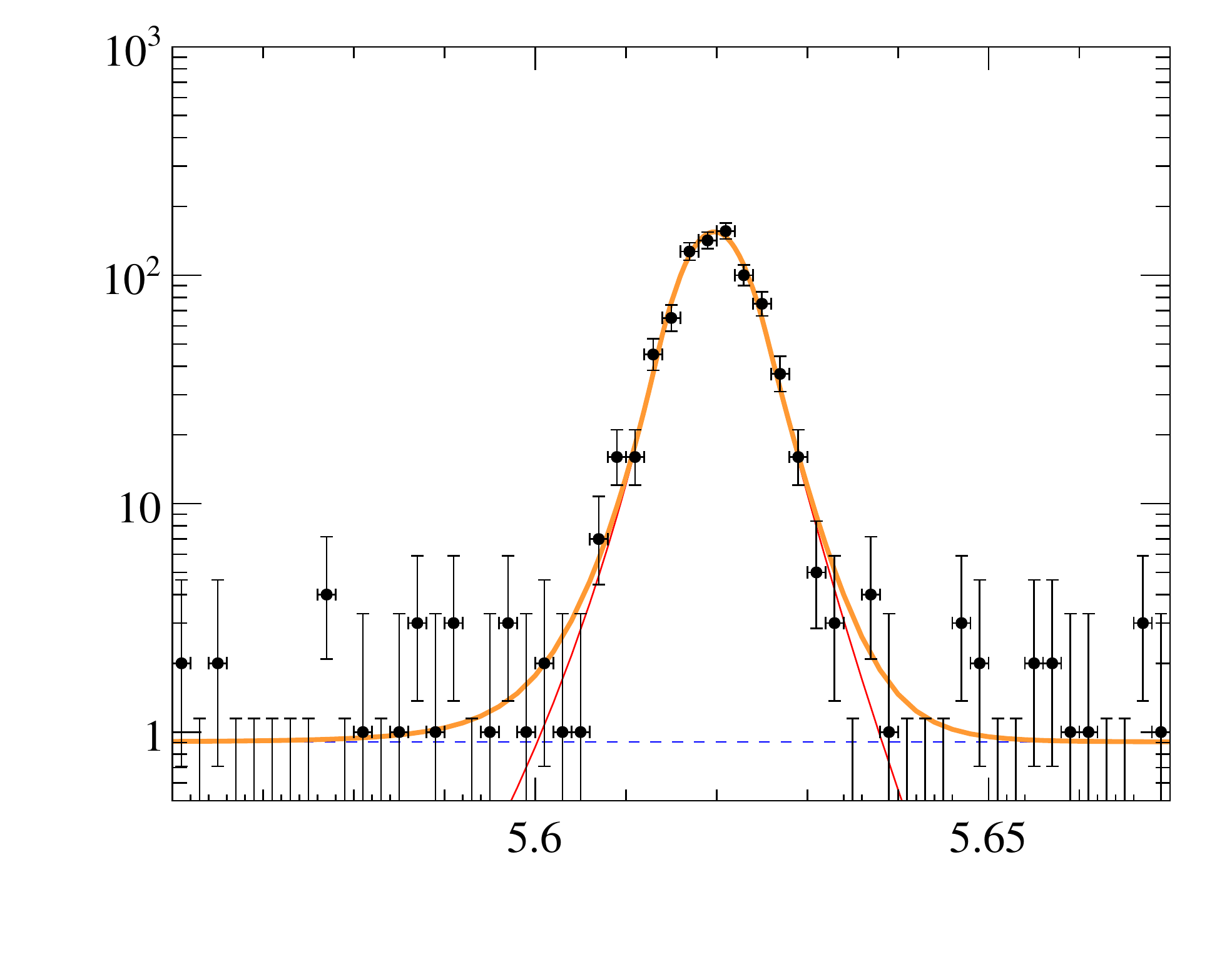}}
    \put(  1,16){\begin{sideways}Candidates/$(5\mevcc)$\end{sideways}}
    \put( 75,16){\begin{sideways}Candidates/$(2\mevcc)$\end{sideways}}
    \put( 32,1.5){$m_{\psitwosPPi}$}
    \put(107,1.5){$m_{\psitwosPK}$}
    \put( 57,1.5){$\left[\!\gevcc\right]$}
    \put(132,1.5){$\left[\!\gevcc\right]$}
    \put( 57,50){\lhcb}
    \put(132,50){\lhcb}
    \put(13,52){\color[rgb]{1,0,0}     {\rule{6mm}{1.0pt}}} 
    \put(13,47){\color[rgb]{0,0,1}     {\hdashrule[0.0ex][x]{6mm}{1.0pt}{1.2mm 0.3mm} } } 
    \put(13,42){\color[rgb]{1,0.6,0.2} {\rule{6mm}{3.0pt}}} 
    \put(20,51){\small{$\Lb\to\psitwos \proton\pim$}}
    \put(20,46){\small{background}}
    \put(20,41){\small{total~fit}} 
    \put(88,52){\color[rgb]{1,0,0}     {\rule{6mm}{1.0pt}}} 
    \put(88,47){\color[rgb]{0,0,1}     {\hdashrule[0.0ex][x]{6mm}{1.0pt}{1.2mm 0.3mm} } } 
    \put(88,42){\color[rgb]{1,0.6,0.2} {\rule{6mm}{3.0pt}}} 
    \put(95,51){\small{$\Lb\to\psitwos \proton\Km$}}
    \put(95,46){\small{background}}
    \put(95,41){\small{total~fit}} 
    \put(14,37){\line(1,0){4}} 
    \put(16,35){\line(0,1){4}} 
    \put(15.5,39){\line(1,0){1}} 
    \put(15.5,35){\line(1,0){1}} 
    \put(14,36.5){\line(0,1){1}} 
    \put(18,36.5){\line(0,1){1}} 
    \put(16,37){\circle*{0.8}}
    \put(20,36){\small{data}}
    \put(89,37){\line(1,0){4}} 
    \put(91,35){\line(0,1){4}} 
    \put(90.5,39){\line(1,0){1}} 
    \put(90.5,35){\line(1,0){1}} 
    \put(89,36.5){\line(0,1){1}} 
    \put(93,36.5){\line(0,1){1}} 
    \put(91,37){\circle*{0.8}}
    \put(95,36){\small{data}}
  \end{picture}
  \caption {\small
    Mass distributions of
    the~(left)~\mbox{\LbpsitwosPPi} and 
    (right)~\mbox{\LbpsitwosPK}~candidates. 
  }
  \label{fig:signal}
\end{figure}

The~mass distributions for the~selected 
\mbox{\LbpsitwosPPi}~and~\mbox{\LbpsitwosPK}~candidates are
shown in Fig.~\ref{fig:signal}.
The~signal yields are determined using unbinned extended 
maximum\nobreakdash-likelihood fits to these distributions.
For~each  distribution the~\Lb~component is described 
by a~modified Gaussian function with power\nobreakdash-law tails on both sides~\cite{Skwarnicki:1986xj,LHCb-PAPER-2011-013}.
The~tail parameters are fixed to values obtained from simulation, and
the~peak position and resolution of the~Gaussian function
are free to vary in the~fit.
The~combinatorial background component is described by
a~monotonic 
second\nobreakdash-order polynomial function with positive curvature.
The~resolution parameters obtained from the~fits are
found to be \mbox{\sigmastud\mevcc} for the~\mbox{\LbpsitwosPPi} channel
and \mbox{\sigmanorm\mevcc} for the~\mbox{\LbpsitwosPK} channel, which
are in good agreement with expectations from simulation.
The~signal yields are determined to be 
\yieldstud~and  \yieldnorm~for   
the~\mbox{\LbpsitwosPPi} and \mbox{\LbpsitwosPK}~decay modes, respectively.

\begin{figure}[tb]
  \setlength{\unitlength}{1mm}
  \centering
  \begin{picture}(150,46)
    \put(  0, 0){\includegraphics*[width=48mm,height=46mm]{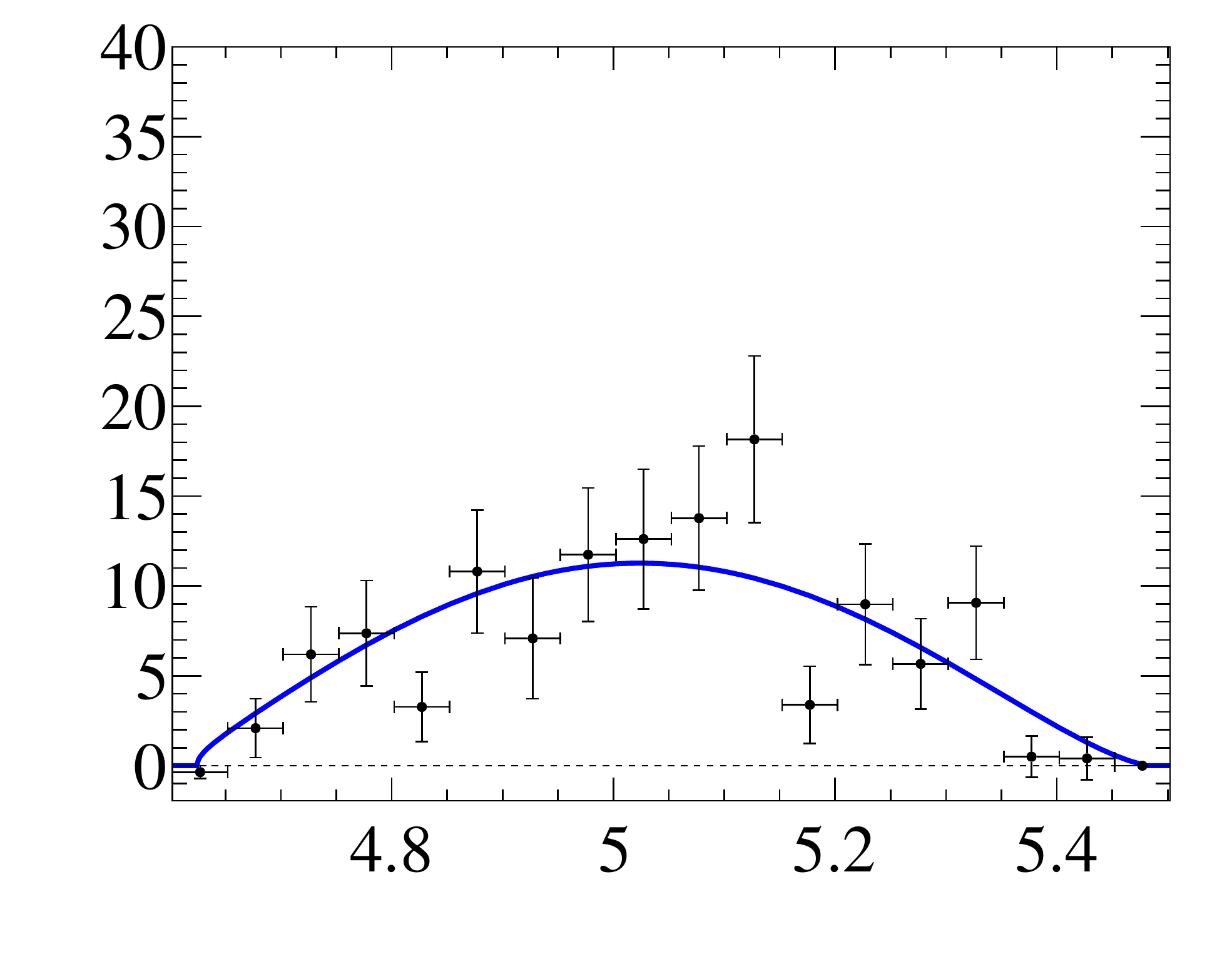}}
    \put( 50, 0){\includegraphics*[width=48mm,height=46mm]{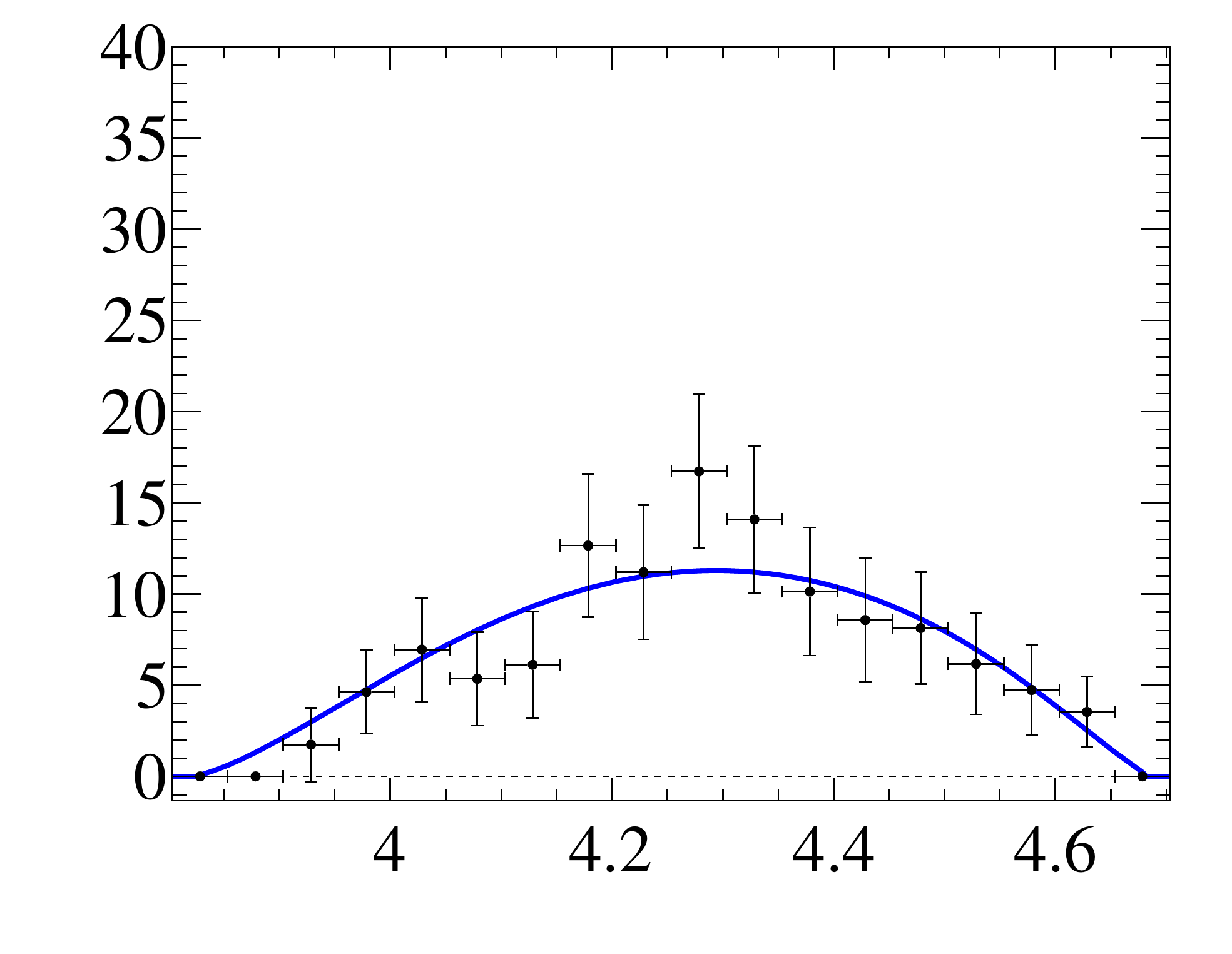}}
    \put(100, 0){\includegraphics*[width=48mm,height=46mm]{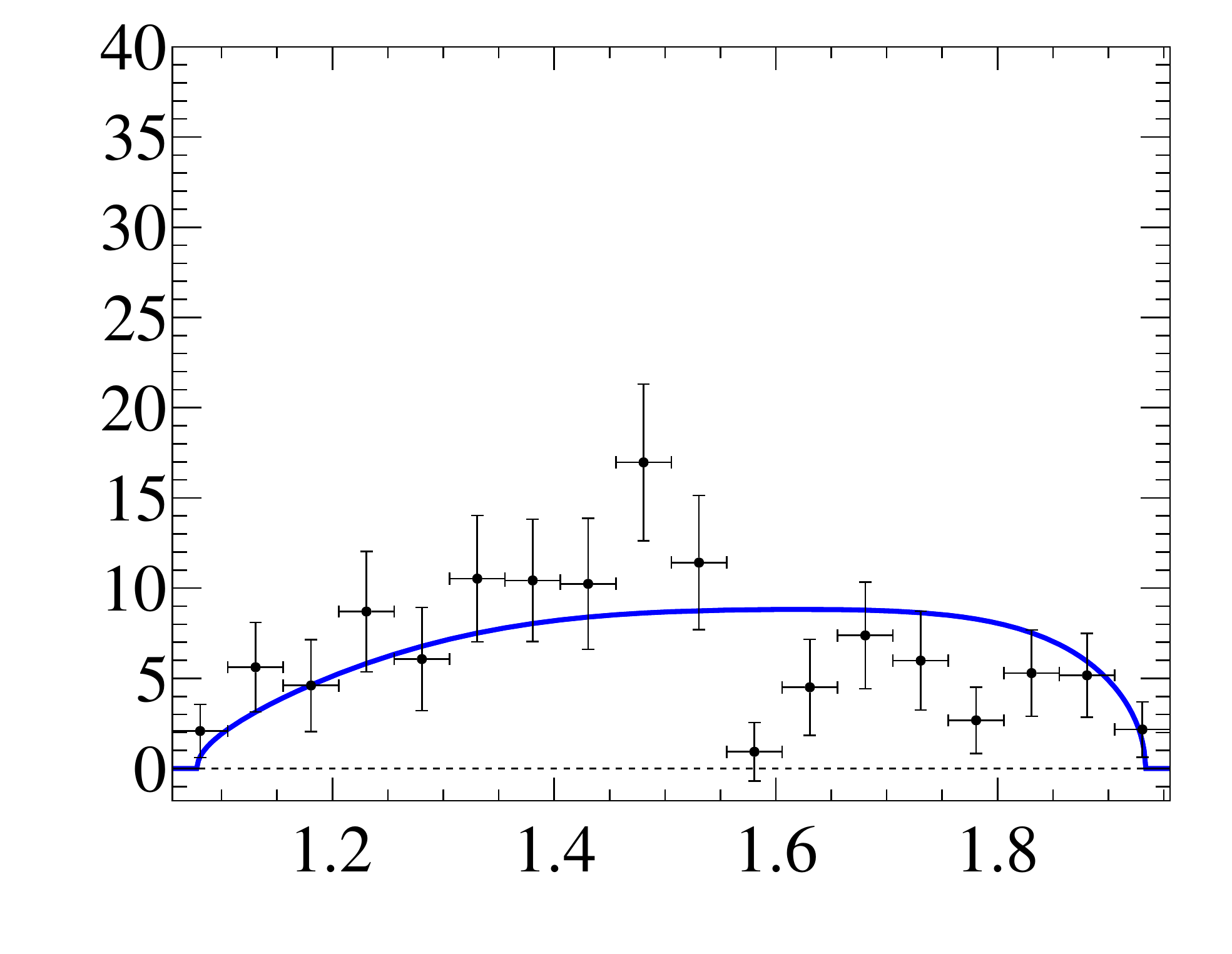}}
    \put( -1.5,15){\small\begin{sideways}$N_{\Lb}/(50\mevcc)$\end{sideways}}
    \put( 48.5,15){\small\begin{sideways}$N_{\Lb}/(50\mevcc)$\end{sideways}}
    \put( 98.5,15){\small\begin{sideways}$N_{\Lb}/(50\mevcc)$\end{sideways}}
    \put( 18  , 0.0){$m_{\psitwosP}$}
    \put( 66  , 0.0){$m_{\psitwosPi}$ }
    \put(121  , 0.0){$m_{\PPi}$ }
    \put( 31.6, 0.0){$\left[\!\gevcc\right]$}
    \put( 81.6, 0.0){$\left[\!\gevcc\right]$}
    \put(131.6, 0.0){$\left[\!\gevcc\right]$}
    \put( 32, 37){\lhcb}
    \put( 82, 37){\lhcb}
    \put(132, 37){\lhcb}
    \put(  9,33){\color[rgb]{0,0,1} {\rule{4mm}{1.5pt}}} 
    \put( 59,33){\color[rgb]{0,0,1} {\rule{4mm}{1.5pt}}} 
    \put(109,33){\color[rgb]{0,0,1} {\rule{4mm}{1.5pt}}} 
    \put( 14,32){\small{simulation}}
    \put( 64,32){\small{simulation}}
    \put(114,32){\small{simulation}}
    \put( 9,37){\line(1,0){4}} 
    \put(11,35){\line(0,1){4}} 
    \put(10.5,39){\line(1,0){1}} 
    \put(10.5,35){\line(1,0){1}} 
    \put( 9,36.5){\line(0,1){1}} 
    \put(13,36.5){\line(0,1){1}} 
    \put(11,37){\circle*{0.7}}
    \put(14,37){\small{data}}
    \put(59,37){\line(1,0){4}} 
    \put(61,35){\line(0,1){4}} 
    \put(60.5,39){\line(1,0){1}} 
    \put(60.5,35){\line(1,0){1}} 
    \put(59,36.5){\line(0,1){1}} 
    \put(63,36.5){\line(0,1){1}} 
    \put(61,37){\circle*{0.7}}
    \put(64,37){\small{data}}
    \put(109,37){\line(1,0){4}} 
    \put(111,35){\line(0,1){4}} 
    \put(110.5,39){\line(1,0){1}} 
    \put(110.5,35){\line(1,0){1}} 
    \put(109,36.5){\line(0,1){1}} 
    \put(113,36.5){\line(0,1){1}} 
    \put(111,37){\circle*{0.7}}
    \put(114,37){\small{data}}
  \end{picture}
  \caption {\small
    Background-subtracted mass distributions of
    the~(left)~\psitwosP,
    (centre)~\psitwosPi~and 
    (right)~\mbox{$\proton\pip$}~combinations
    in the~\mbox{\LbpsitwosPPi} decay compared
    with distributions obtained from a~phase\nobreakdash-space simulation.
  }\label{fig:rest}
\end{figure}
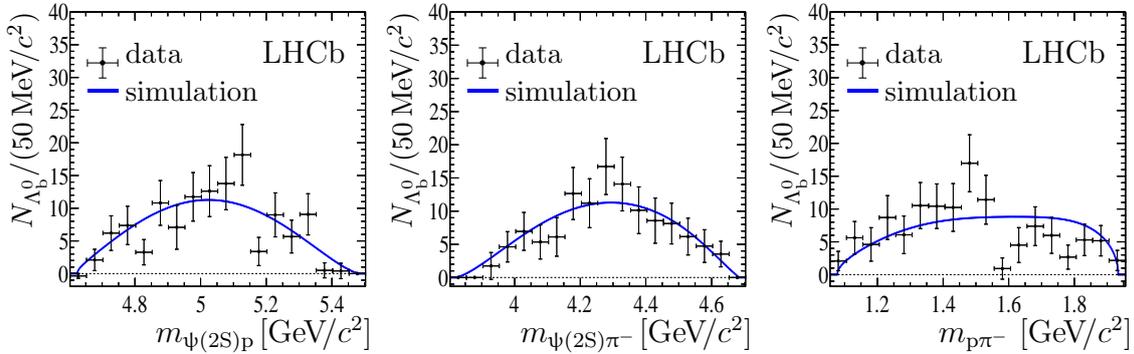

The~resonance structure of the~\mbox{\LbpsitwosPPi} decay is investigated
using the~\sPlot~technique~\cite{Pivk:2004ty} for background subtraction, 
with the~reconstructed \mbox{\psitwosPPi} mass
as the~discriminating variable.
The~background\nobreakdash-subtracted mass distributions of
\psitwosP, \psitwosPi~and \PPi~combinations
are shown in Fig.~\ref{fig:rest}, along with those obtained
from simulated decays generated according to a~phase\nobreakdash-space model.
The~\psitwosP~and \psitwosPi~mass distributions 
show no evidence for contributions from exotic states.   
The~mass distribution of the~\PPi~combination 
differs from the~phase\nobreakdash-space model, 
indicating possible contributions from 
excited $\mathrm{N}^0$ and $\Delta^0$~states.
Further studies with a~larger data sample will provide a~deeper insight into
the~underlying structure of the~\mbox{\LbpsitwosPPi} decay.

The~ratio of branching fractions $R_{\pion/\kaon}$, defined in Eq.\,\eqref{eq:br_ratio},
is measured as
\begin{equation}
  R_{\pion/\kaon} = \dfrac{N_{\Lb\to\psitwosPPi}}{N_{\Lb\to\psitwosPK}}\, 
  \dfrac{\upvarepsilon_{\Lb\to\psitwosPK}}{\upvarepsilon_{\Lb\to\psitwosPPi}}\,,
\end{equation}
where $N$~represents the~measured yield and $\upvarepsilon$~denotes
the~efficiency of the~corresponding decay.
The~efficiency is defined as the~product of the~geometric acceptance and
the~detection, reconstruction, selection and trigger efficiencies.
The~hadron\nobreakdash-identification efficiencies as functions 
of kinematics and the~event multiplicity are determined from data
using the~following calibration samples of 
low\nobreakdash-background decays:
\mbox{$\decay{\Dstarp}{\Dz(\to{\Km\pip})\pip}$},
\mbox{$\KS\to\pip\pim$} and \mbox{$\Ds\to\Pphi(\to\Kp\Km)\pip$}
for kaons and pions; and
\mbox{$\Lz\to\proton\pim$} and 
\mbox{$\Lc\to\proton\Kp\pim$}
for protons~\cite{LHCb-DP-2012-003,LHCb-DP-2018-001}.
The~remaining efficiencies are determined using simulation.
The~$\pt$~and rapidity spectra and 
the~lifetime of the~\Lb~baryons in simulated samples 
are adjusted to match those observed in a~high\nobreakdash-yield 
low\nobreakdash-background sample of 
reconstructed \mbox{$\Lb\to\jpsi\proton\Km$}~decays.
The~simulated samples 
are produced according to a~phase\nobreakdash-space decay model.
The~simulated~\LbpsitwosPK~decays are corrected to reproduce 
the~$\proton\Km$~mass and $\cos \uptheta_{\proton\Km} $~distributions observed in data,
where $\uptheta_{\proton\Km}$ is the~helicity angle of the~$\proton\Km$~system,
defined as the~angle between the~momentum vectors of the~kaon and \Lb~baryon
in the~$\proton\Km$~rest frame.
To~account for imperfections in the~simulation of charged tracks,
corrections obtained using data\nobreakdash-driven techniques are 
also applied~\cite{LHCb-DP-2013-002}.

The~efficiencies are determined separately for each data\nobreakdash-taking 
period and are combined according to the~corresponding 
luminosity~\cite{LHCb-PAPER-2014-047} 
for each period and the~known 
production cross\nobreakdash-section of 
$\bquark\bquarkbar$~pairs 
in the~LHCb acceptance~\cite{LHCb-PAPER-2010-002,
  LHCb-PAPER-2011-003,
  LHCb-PAPER-2013-016,
  LHCb-PAPER-2015-037,
  LHCb-PAPER-2016-031}.
The~ratio of the~total~efficiency of the~normalization
channel to that of the~signal channel
is determined to be
\begin{equation}
  \dfrac{\upvarepsilon_{\LbpsitwosPK}}{\upvarepsilon_{\LbpsitwosPPi}} = \effratio\,,
  \label{eq:eff_ratio}
\end{equation}
where only the~uncertainty that arises from the~sizes of the~simulated samples is given. 
Additional sources  of uncertainty are discussed in the~following section.
The~kaon identification efficiency, entering into $\upvarepsilon_{\LbpsitwosPK}$, 
is the~main factor causing non\nobreakdash-equality of the~total efficiencies for 
the~signal and normalization channels.

\section{Systematic uncertainties}
\label{sec:systematics}

Since the~signal and normalization decay channels have similar kinematics 
and topologies, most systematic uncertainties cancel in 
the~ratio $R_{\pion/\kaon}$, 
\eg those related to muon identification.
The~remaining contributions to the~systematic uncertainty
are listed in Table~\ref{tab:systematics} and discussed below.

To~estimate the~systematic uncertainty related to the~fit model, 
pseudoexperiments are sampled from the~baseline fit models with 
all parameters fixed from those obtained from the fits to the~data.
For~each pseudoexperiment fits are performed 
with a~number of alternative models for the~signal and background components
and the~ratio $R_{\pion/\kaon}$ is computed.
A~generalized Student's t-distribution~\cite{Jackman} and 
an~Apollonios function~\cite{Santos:2013gra} are used as~alternative 
models for the~signal component,
while 
polynomial functions of the~second and the~third order 
with various constraints for monotonicity and convexity are used
as alternative backgrounds.
The~maximum relative bias found  for $R_{\pion/\kaon}$ is~\systFitModel{\%},
which is assigned as a~relative systematic uncertainty.

The~uncertainty related to the~imperfect knowledge of 
the~\Lb~decay model used for the~simulation of 
the~\mbox{\LbpsitwosPK}~decays is estimated by varying  
the~correction factors obtained from 
kinematic distributions observed in data.  
Changing these~correction factors within 
their~statistical uncertainties 
causes a~negligible variation of 
the~efficiency~$\upvarepsilon_{\LbpsitwosPK}$. 
For~the~\mbox{\LbpsitwosPPi}~signal decays the~observed 
two\nobreakdash-body mass distributions 
are in agreement with the~phase\nobreakdash-space model
used in the~simulation. The~corresponding uncertainty due to 
the~unknown decay kinematics of the~\mbox{\LbpsitwosPPi}~signal decays 
is small and therefore neglected. 

An~additional uncertainty arises from the~differences between data and simulation, 
in particular those affecting the~efficiency for 
the~reconstruction of charged\nobreakdash-particle tracks. 
The~small difference in the~track\nobreakdash-finding 
efficiency between data and simulation is corrected using 
a~data\nobreakdash-driven technique~\cite{LHCb-DP-2013-002}.
The~uncertainties in these~correction factors together with 
the~uncertainties in the~hadron\nobreakdash-identification efficiencies,  
related to the~finite size of the~calibration 
samples~\cite{LHCb-DP-2012-003,LHCb-DP-2018-001},
are propagated  to the~ratio of total efficiencies by means of 
pseudoexperiments. 
This~results in a~systematic uncertainty of \systPID{\%} 
associated with the~track reconstruction and hadron identification.

The~systematic uncertainty on the~efficiency of the~trigger has been previously
studied using high\nobreakdash-yield \mbox{$\decay{\Bp}{\jpsi\Kp}$}~and
\mbox{$\decay{\Bp}{\psitwos\Kp}$}~decays by comparing ratios of trigger efficiencies
in data and simulation~\cite{LHCb-PAPER-2012-010}.
Based on these comparisons a~relative uncertainty of \mbox{\systTrigger{\%}} is assigned.

Another source of uncertainty is the~potential disagreement between data and simulation
in the~estimation of efficiencies, due to effects not considered above. 
This is studied using a~high\nobreakdash-yield low\nobreakdash-background 
sample of \mbox{\LbjpsiPK}~decays, by varying the~selection criteria in ranges 
that lead to as much as $\pm20\%$ differences in the~measured signal yields. 
The~resulting variations in the~efficiency\nobreakdash-corrected yields 
do not exceed $1\%$~for  all inspected selection criteria. 
The~value of $1\%$~is taken as a~corresponding systematic uncertainty.

Finally, the~\mbox{\systSSS{\%}}~relative uncertainty 
in the~ratio of efficiencies from~Eq.\,\eqref{eq:eff_ratio} 
is assigned as a~systematic uncertainty due to the~finite size of the~simulated samples.

\begin{table}[t]
  \centering
  \caption{\small
    Relative systematic uncertainties for the~ratio of branching fractions.
    The~total uncertainty is the~quadratic sum of the~individual contributions.
  }\label{tab:systematics}
  \vspace*{3mm}
  \begin{tabular*}{0.80\textwidth}{@{\hspace{3mm}}l@{\extracolsep{\fill}}c@{\hspace{3mm}}}
    Source   		& Uncertainty~$\left[\%\right]$	
    \\[1mm]
    \hline
    \\[-3mm]
    Fit model				& \systFitModel		\\
    Track reconstruction and hadron identification & \systPID			\\
    Trigger				& \systTrigger		\\
    Selection criteria			& \systDaSiAgr		\\
    Size of the~simulation samples      & \systSSS	        
    \\[1mm]
    \hline
    \\[-3mm]
    Total				& \systTotal
  \end{tabular*}
\end{table}

\section{Results and summary}
\label{sec:Result}

The~Cabibbo\nobreakdash-suppressed decay~\mbox{\LbpsitwosPPi}~is observed
using a~data sample 
collected by the~\lhcb
experiment in proton\nobreakdash-proton collisions 
corresponding to 1.0, 2.0 and 1.9\invfb of  integrated
luminosity at 
centre\nobreakdash-of\nobreakdash-mass
energies of 7, 8 and 13\tev, respectively.
The~observed yield of \mbox{\LbpsitwosPPi}~decays is \yieldstud. 
Using the~\LbpsitwosPK decay as a~normalization channel, the~ratio of
the~branching fractions is measured to be
\begin{equation*}
  R_{\Ppi/\kaon} = 
  \dfrac{\BR\left(\LbpsitwosPPi\right)}{\BR\left(\LbpsitwosPK\right)}=\brratio\,,
\end{equation*}
where the~first uncertainty is statistical and the~second is systematic.
Neglecting the~resonance structures in the~\LbpsitwosPPi and \LbpsitwosPK~decays, 
the~calculated value for the~ratio $R_{\Ppi/\kaon}$ is 
\begin{equation*}
  R^{\mathrm{th}}_{\Ppi/\kaon} \approx 
  \dfrac{ \Phi_3\left(\LbpsitwosPPi\right)}
  { \Phi_3\left(\LbpsitwosPK\right)} \times
  \tan^2 \uptheta_{\mathrm{C}} \simeq 11\%\,,
\end{equation*}
where $\Phi_3$ denotes the~full three\nobreakdash-body phase\nobreakdash-space and 
$\uptheta_{\mathrm{C}}$ is the~Cabibbo angle~\cite{Cabibbo:1963yz}. 
The~measured value is in a~good agreement with this estimate.

The~branching fraction $\BR\left(\LbpsitwosPPi\right)$ is  calculated 
using the value of $\BR\left(\LbpsitwosPK\right)=\left(6.29 \pm 0.23 \pm 0.14 ^{+1.14}_{-0.90}\right)\times 10^{-5}$~\cite{LHCb-PAPER-2015-060} as 
\begin{equation*}
  \BR\left(\LbpsitwosPPi\right)= 
  \left( 7.17 \pm 0.82 \pm 0.33 ^{+1.30}_{-1.03}\right) \times 10^{-6}\,,
\end{equation*}
where the~first uncertainty is statistical, 
the~second systematic\,(including the~statistical and systematic uncertainties from 
$\BR\left(\LbpsitwosPK\right)$) and the~third arises from the~uncertainties 
in the~branching fractions of 
the~\mbox{$\Lb\to\jpsi\proton\Km$}, 
\mbox{$\psitwos\to\jpsi\pip\pim$},
\mbox{$\psitwos\to\mathrm{e}^{+}\mathrm{e}^{-}$} and 
\mbox{$\jpsi\to\mathrm{e}^{+}\mathrm{e}^{-}$}~decays~\cite{PDG2018}.

The~$\psitwos\proton$ and $\psitwos\pim$~mass spectra are investigated
and no~evidence for contributions from exotic states is found. 
With a~larger data sample 
a~detailed amplitude analysis of this decay could  be performed,
making it possible  to search for small contributions from exotic states.

\section*{Acknowledgements}
%
%
\noindent We~express our gratitude to our colleagues in the~CERN
accelerator departments for the~excellent performance of the~LHC. 
We~thank the~technical and administrative staff at the~LHCb
institutes. 
We~acknowledge support from CERN and from the~national
agencies: CAPES, CNPq, FAPERJ and FINEP\,(Brazil); 
MOST and NSFC\,(China); 
CNRS/IN2P3\,(France); 
BMBF, DFG and MPG\,(Germany); 
INFN\,(Italy); 
NWO\,(Netherlands); 
MNiSW and NCN\,(Poland); 
MEN/IFA\,(Romania); 
MinES and FASO\,(Russia); 
MinECo\,(Spain); 
SNSF and SER\,(Switzerland); 
NASU\,(Ukraine); 
STFC\,(United Kingdom); 
NSF\,(USA).  
We~acknowledge the~computing resources that are provided by CERN, 
IN2P3\,(France), 
KIT and DESY\,(Germany), 
INFN\,(Italy), 
SURF\,(Netherlands),
PIC\,(Spain), 
GridPP\,(United Kingdom), 
RRCKI and Yandex~LLC\,(Russia), 
CSCS\,(Switzerland), 
IFIN\nobreakdash-HH\,(Romania), 
CBPF\,(Brazil),
PL\nobreakdash-GRID\,(Poland) and 
OSC\,(USA). 
We~are indebted to the~communities
behind the multiple open\nobreakdash-source software packages on which we depend.
Individual groups or members have received support from 
AvH Foundation\,(Germany), 
EPLANET, Marie Sk\l{}odowska\nobreakdash-Curie Actions and ERC\,(European Union), 
ANR, Labex P2IO and OCEVU, and R\'{e}gion Auvergne\nobreakdash-Rh\^{o}ne\nobreakdash-Alpes\,(France), 
Key Research Program of Frontier Sciences of CAS, CAS PIFI, and the~Thousand Talents Program\,(China),
RFBR, RSF and Yandex~LLC\,(Russia), 
GVA, XuntaGal and GENCAT\,(Spain),
Herchel Smith Fund, the~Royal Society, the~English\nobreakdash-Speaking Union and
the~Leverhulme Trust\,(United Kingdom).

\clearpage

\addcontentsline{toc}{section}{References}
\setboolean{inbibliography}{true}
\bibliographystyle{LHCb}
\bibliography{local,main,LHCb-PAPER,LHCb-CONF,LHCb-DP,LHCb-TDR}

\newpage
\centerline{\large\bf LHCb collaboration}
\begin{flushleft}
\small
R.~Aaij$^{27}$,
B.~Adeva$^{41}$,
M.~Adinolfi$^{48}$,
C.A.~Aidala$^{73}$,
Z.~Ajaltouni$^{5}$,
S.~Akar$^{59}$,
P.~Albicocco$^{18}$,
J.~Albrecht$^{10}$,
F.~Alessio$^{42}$,
M.~Alexander$^{53}$
% ...

{\footnotesize \it
% ...
$ ^{5}$Clermont Universit{\'e}, Universit{\'e} Blaise Pascal, CNRS/IN2P3, LPC, Clermont-Ferrand, France\\
% ...
$ ^{10}$Fakult{\"a}t Physik, Technische Universit{\"a}t Dortmund, Dortmund, Germany\\
% ...
$ ^{18}$INFN Laboratori Nazionali di Frascati, Frascati, Italy\\
% ...
$ ^{27}$Nikhef National Institute for Subatomic Physics, Amsterdam, Netherlands\\
% ...
$ ^{41}$Instituto Galego de F{\'\i}sica de Altas Enerx{\'\i}as (IGFAE), Universidade de Santiago de Compostela, Santiago de Compostela, Spain\\
$ ^{42}$European Organization for Nuclear Research (CERN), Geneva, Switzerland\\
% ...
$ ^{48}$H.H. Wills Physics Laboratory, University of Bristol, Bristol, United Kingdom\\
% ...
$ ^{53}$School of Physics and Astronomy, University of Glasgow, Glasgow, United Kingdom\\
% ...
$ ^{59}$University of Cincinnati, Cincinnati, OH, United States\\
% ...
$ ^{61}$Syracuse University, Syracuse, NY, United States\\
% ...
$ ^{73}$University of Michigan, Ann Arbor, United States, associated to $^{61}$\\
% ...
}
\end{flushleft}

\end{document}